\documentclass[12pt]{iopart}
\usepackage{iopams}
\usepackage{indentfirst}
\usepackage{braket}
\usepackage{makeidx}
\usepackage{slashed}
\usepackage[dvipdfmx]{graphicx}
\usepackage{xcolor}
\usepackage[normalem]{ulem}

\newcommand\subHalmos{\protect\rule[0.5pt]{3pt}{3pt}}

\begin{document}
\title{Effective Floquet-Gibbs states for  dissipative quantum systems}
\author{Tatsuhiko Shirai$^1$, Juzar Thingna$^{2,3}$, Takashi Mori$^1$, Sergey Denisov$^{2,4}$, Peter H{\"a}nggi$^{2,3,4,5}$, Seiji Miyashita$^1$}
\address{%
$^1$Department of Physics, Graduate School of Science, The University of Tokyo, 7-3-1 Hongo, Bunkyo-Ku, Tokyo 113-8656, Japan}
\address{%
$^2$Institute of Physics, University of Augsburg, Universit\"atstra{\ss}e 1, D-86135 Augsburg, Germany}
\address{%
$^3$Nanosystems Initiative Munich, Schellingstra{\ss}e 4, D-80799 M\"unchen, Germany}
\address{%
$^4$Department of Applied Mathematics, Lobachevsky State University of Nizhny Novgorod, Nizhny Novgorod 603950, Russia}
\address{%
$^5$Department of Physics, National University of Singapore, 117551 Singapore, Republic of Singapore }
\ead{\mailto{shirai@spin.phys.s.u-tokyo.ac.jp}}
%Present address: Singapore-MIT Alliance for Research and Technology (SMART) Centre, Singapore 138602

\begin{abstract}
A periodically driven quantum system, when coupled to a heat bath,
relaxes to a non-equilibrium asymptotic state. In the general situation, the retrieval of
this asymptotic state presents a rather non-trivial task.
It was recently shown  that in the limit of an infinitesimal coupling, using
the so-called rotating wave approximation (RWA),
and under strict conditions imposed on the time-dependent system Hamiltonian,
the asymptotic state can attain the Gibbs form.
A Floquet-Gibbs state is characterized by a density matrix which is diagonal in the Floquet basis of the
system Hamiltonian with the diagonal elements obeying a Gibbs distribution, being
parametrized by the corresponding Floquet quasi-energies. Addressing the non-adiabatic driving regime, upon using the Magnus expansion, we employ the concept of a corresponding effective Floquet Hamiltonian. In doing so we go beyond the conventionally used RWA and demonstrate
that the idea of  Floquet-Gibbs states can be extended to the realistic case of a weak, although finite system-bath coupling, herein termed  \emph{effective} Floquet-Gibbs states.
\end{abstract}

\maketitle
\section{Introduction}
\label{intro}
When  coupled to a heat bath, a quantum system with a time-\textit{independent}
Hamiltonian typically relaxes to an overall thermal equilibrium state~\cite{nakajima1958quantum, zwanzig1960ensemble,hanggi2005chaos}.
In the infinitesimal coupling limit, this thermal state is specified
by the canonical Gibbs density matrix $\varrho$; i.e.,  $\varrho \propto e^{-\beta H_{\rm S}}$, where $H_{\rm S}$
is the system Hamiltonian and $\beta$ denotes the inverse temperature
of the heat bath~\cite{feynman1998statistical}. Although this result is quite intuitive,
the mechanism behind its universal form and
its emergence from the system-specific quantum evolution remains the focus
of active studies and debates up to this date~\cite{spohn1977algebraic,
tasaki1998quantum, reimann2007typicality, mori2008dynamics, thingna2012generalized}.

The case of a \textit{periodically modulated} quantum system coupled to a bath is even more challenging.
No universal closed-form expression for the asymptotic state
is known and hence one has to analyze the specific dynamics
of the system of interest, analytically or numerically. When isolated, i.e. when the system is not coupled to a heat bath, periodically
driven quantum systems have been extensively studied and a variety of intriguing phenomena have been discovered.
On the single-particle level,  prominent  effects~\cite{grifoni1998driven, kayanuma2008coherent}
such as dynamical localization~\cite{dunlap1986dynamic}
and the coherent destruction of tunneling phenomenon~\cite{grossmann1991coherent, grossmann1991tunneling}
have been discovered theoretically and have become validated in experiments. Moreover, the use of periodic driving fields has found applications recently
to create new phases or topological band structures which otherwise are absent in equilibrium
\cite{eckardt2005superfluid, lindner2011floquet, zenesini2009coherent, wang2013observation}.

Floquet states of many-body systems are presently actively explored  by using
 the idea of a time-\textit{in}dependent \textit{effective} Hamiltonian (sometimes also termed  `Floquet Hamiltonian'),
whose eigenstates approximately coincide with the Floquet states
of the original time-dependent Hamiltonian at stroboscopic instants of multiple periods of the underlying high-frequency periodic drive;
see the recent review by Bukov \textit{et al.}~\cite{bukov2015universal}.

Clearly, the universal Gibbs form of the density matrix comprises a strong appeal, so it is not surprising
that several attempts have been made to generalize the idea to the case of periodically
driven open quantum systems. For example, such a Gibbs form arises
in the model of a single particle subjected to a  modulated
harmonic potential~\cite{breuer2000quasistationary, langemeyer2014energy}
and systems in which the time dependence of the Hamiltonian
can be eliminated through a unitary transformation~\cite{grifoni1998driven, iadecola2013generalized}.
In case of non-integrable quantum systems,
it is very tempting to use Floquet states as the eigenbasis to
write down the density matrix and then search for the limit where
the latter acquires the Gibbs form. But before doing so, one should
make a guess as to what quantity should take the place for the energy $\tilde{E}_i$
in the diagonal elements, $\varrho_{ii} \propto e^{-\beta\tilde{E}_i}$.
A very natural idea that this role could be played by the average energy of the $i$-th Floquet state;
i.e., the expectation value of the Hamiltonian averaged over one period of the driving,
has been tested in Refs.~\cite{breuer2000quasistationary, ketzmerick2010statistical}.
The results demonstrated that although the density matrices have
diagonal elements reasonably close to the Boltzmann factors
(though with an `effective temperature' different from the actual temperature of the heat bath),
there is also a tangible deviation from the Gibbs form. This difference was
related to the coexistence of (semi-classically) chaotic and regular
Floquet states~\cite{ketzmerick2010statistical}.

Recently, some of the authors presented an alternative idea,
upon introducing the notion of the Floquet-Gibbs states~\cite{shirai2015condition},
i.e. the states whose density matrices have Gibbs form in the Floquet basis of the system Hamiltonian,
with $\tilde{E}_i$ being the \textit{quasi}-energy of the corresponding Floquet state.

The main problem of using quasi-energies as effective energies $\tilde{E}_i$ is that they are
mere phase factors and thus are defined up to multiples of $\hbar \Omega$, where $\Omega$ is the driving frequency ~\cite{grifoni1998driven}.
Because of this ambiguity, quasi-energies have  little to do with the actual energies
of the corresponding Floquet states. For example, two states of very different energies can come close to each other
inside the first Brillouin zone $[- \hbar \Omega/2, \hbar \Omega/2)$ and produce an avoided crossing which leads
to  resonance effects~\cite{timber, weitz2016}.
In order to illustrate the conditions required to meet a Floquet-Gibbs state let us consider a time-dependent
Hamiltonian of the type $H_{\rm S}(t) = H_0 + \xi H_{\rm ex}(t)$ with a time-periodic Hermitian operator
$H_{\rm ex}(t+T) = H_{\rm ex}(t)$, $T = 2\pi/\Omega$, and $\xi$ being the control parameter to tune
the strength of the modulation. Then, the ambiguity related to the quasi-energies disappears when the condition --
\begin{itemize}
\item[] (i) the angular driving frequency $\Omega$ has to be much larger than the
spectral width of the system Hamiltonian $H_0$
\end{itemize}
-- is obeyed. This condition implies that the energy spectrum of the system
fits into the first Brillouin zone and the natural way of ordering the quasi-energies exists.
Technically, this condition then requires fast driving.

In order to have the asymptotic density matrix diagonal in the Floquet basis,
it also has to be guaranteed that all dissipative effects are relevant on a time
scale  larger than \textit{any} intrinsic characteristic timescale of the
isolated system (including the period of the driving $T$).
Under this condition the evolution of the off-diagonal elements of the system density matrix becomes decoupled
from the evolution of the diagonal elements so that the former decay exponentially fast  in time.
This amounts to a so-called `rotating-wave approximation' (RWA)~\cite{breuer2000quasistationary, langemeyer2014energy,
ketzmerick2010statistical, kohn2001periodic, blumel1991dynamical, oelschlagel1993a, oelschlagel1993b, kohler1997floquet, Dehghani2014dissipative}. Because the relaxation time is inversely proportional to the square
of the system-bath coupling $\lambda$, $\tau_{\rm relax} \propto \lambda^{-2}$,
the RWA is only valid when the system-bath coupling becomes infinitesimal.

Finally, there are two further specific conditions that ensure the asymptotic states to be
the Floquet-Gibbs state~\cite{shirai2015condition}:
\begin{itemize}
\item[]
(ii) the time-dependent part of the system Hamiltonian, $H_{\rm ex}(t)$,
should commute with itself at different instants of time, $[H_{\rm ex}(t_1),H_{\rm ex}(t_2)] = 0$;
\item[]
(iii) the time-dependent part of the system Hamiltonian, $H_{\rm ex}(t)$, and the system-bath
interaction Hamiltonian should commute.
\end{itemize}

The three conditions, (i-iii), therefore limit the class of suitable physical models.
Namely, condition (i) either restricts the system Hilbert space to
relatively small sizes \textit{or} confines allowed modulations to the high-frequency
limit (where their effect is reduced to a mere re-normalization of the stationary Hamiltonian). In contrast,
condition (ii) is less serious; it is fulfilled with the choice $H_{\rm ex}(t) = f(t)H_{1}$, where $f(t)$ is a time-periodic
scalar function and $H_1$ is a system Hermitian operator. Most of the currently used
models belong to
this class~\cite{bukov2015universal, lazarides2014equilibrium, dalessio2014long-time, ponte2015periodically, zhou2015thermo}.
Condition (iii) requires fine-tuning of system-bath coupling operators, which is difficult to control.

With this work we  address the following two questions:
\begin{itemize}
\item
\textit{Can the concept of Floquet-Gibbs states be
extended to the case of a weak but \textit{finite} system-bath coupling?}
\item
\textit{Can some of the conditions } (i-iii) \textit{ be relaxed?}
\end{itemize}

Assume that condition (i) is broken, but the system is subjected to high frequency periodic modulation. In this case the heating rate of the system, $\tau_{\rm heat}$ (i.e., the speed with which the modulations pump energy into the system~\cite{bunin2011}), is extremely large~\cite{kuwahara2016floquet, mori2016rigorous, abanin2015asymptotic, abanin2015effective}.
Thus, when the system couples to the heat bath at weak but finite strength, the relaxation rate $\tau_{\rm relax}^{-1}$ might be larger than the heating rate $\tau_{\rm heat}^{-1}$, i.e., $\tau_{\rm heat} > \tau_{\rm relax}$, which substantially changes the asymptotic state (see Sec.~\ref{condition} for more details). Also in this case the RWA is no longer applicable (since $\tau_{\rm relax}$ is not the longest timescale) and there is no guarantee that it will describe an approximately valid asymptotic state.

In order to go beyond the limit imposed by the condition (i) we need to avoid
the ambiguity in the definition of effective energies and find a suitable candidate
to test the hypothesis of effective Floquet-Gibbs states. The quasi-energies are not suitable
for this purpose because there is no unique way to extract the `correct'
effective Hamiltonian. This is because the logarithm of the Floquet unitary operator possesses many branches. The first intuitive candidate for a time-independent Floquet Hamiltonian is obtained by summing up
the Magnus expansion of the original system Hamiltonian $H_{\rm S}(t)$~\cite{bukov2015universal}.
However, when it comes to a technical realization
of this idea, one faces a problem: the expansion simply does not converge~\cite{blanes2009magnus}. This recently became an issue of active
research in the field
of many-body quantum physics~\cite{kuwahara2016floquet, mori2016rigorous, abanin2015asymptotic, abanin2015effective}.
There was a proposal to circumvent this obstacle by truncating the Magnus series and it was claimed that such an approximation could accurately describe the long-lived transient
states~\cite{kuwahara2016floquet, mori2016rigorous, abanin2015asymptotic, abanin2015effective, mori2015floquet}.
In this situation one not only obtains uniquely defined `effective energies', that are eigenvalues of the
truncated Floquet Hamiltonian, but also the eigenbasis is more preferable
than the Floquet basis to express the asymptotic non-equilibrium state.

Condition (iii) becomes irrelevant when the response of the bath to the high frequency driving field is weak, e.g., the frequency of the driving field is much larger than the cutoff frequency of the bath spectral density $\omega_c$.
Here $\omega_c$ represents a characteristic energy scale of the system-bath coupling (see Sec.~\ref{condition} for more details). In conclusion, by relaxing conditions (i) and (iii) we can broaden the subclass
of systems for which an analog of the Gibbs distribution can indeed be
introduced. We illustrate this conjecture by using a non-integrable  spin chain model driven by a time-periodic magnetic field.

The work is organized as follows. In Sec.~\ref{redfield} we present an overview of the
Redfield formalism and outline the theoretical basis of the problem. In Sec.~\ref{Floquet-Gibbs}
we define the effective Floquet-Gibbs state, and in Sec.~\ref{condition} we explain our conjectures for its emergence. Sec.~\ref{spinchain} introduces the non-integrable spin
chain model. In Sec.~\ref{finitecoupling}, by using the spin chain model as a testbed,
we demonstrate how condition (i) can be lifted via a finite system-bath coupling.
In Sec.~\ref{timescale}, we show that the condition (iii) is not necessary when $\omega_c \ll \Omega$. We conclude with a discussion of open issues in Sec.~\ref{conclusion}.

\section{Periodic asymptotic states via the Redfield equation}
\label{redfield}
We start with the total Hamiltonian,
\begin{equation}
\label{eq:Ham}
H(t)=H_{\rm S}(t)+H_{\rm B}+\lambda H_{\rm I}, \quad H_{\rm I}=\sum_{\alpha} X^{\alpha} \otimes Y^{\alpha},
\end{equation}
where $H_{\rm S}(t)$ and $H_{\rm B}$ denote the Hamiltonian
of the system and the bath. $H_{\rm I}$ is the interaction Hamiltonian, and $\lambda$ is a dimensionless parameter indicating the strength of system-bath coupling.
We set $X^{\alpha}$ and $Y^{\alpha}$ to be Hermitian operators in the Hilbert space of the system and bath respectively.
The system Hamiltonian is time-periodic, $H_{\rm S}(t)=H_{\rm S}(t+T)$,
where $T$ is the period of the driving.
Within the Born-Markov approximation, the reduced density matrix of
the system $\rho(t)$ obeys the Redfield equation~\cite{kubo1991statistical, petruccione2002theory}:
\begin{eqnarray}
\frac{d\rho (t)}{dt}=&-\frac{i}{\hbar} [H_{\rm S}(t), \rho (t)] \nonumber\\
&-\frac{\lambda^2}{\hbar^2} \sum_{\alpha} \int_0^{\infty}
\Big\{ \langle Y^{\alpha}(\tau ) Y^{\alpha} \rangle _{\beta}[ X^\alpha, X^{\alpha}(t, t-\tau) \rho (t) ]\nonumber\\
& \qquad\qquad\qquad -\langle Y^{\alpha} Y^{\alpha} (\tau ) \rangle_{\beta} [ X^{\alpha}, \rho (t) X^{\alpha}(t, t-\tau ) ] \Big\} d\tau,
\label{eq:QMEop}
\end{eqnarray}
where $X^{\alpha}(t', t)=U(t', t) X^{\alpha} U^{\dagger} (t', t)$
with $U(t', t)={\cal T} e^{ - \frac{i}{\hbar} \int_{t}^{t'} H_{\rm S}(\tau ) d\tau}$,
and $Y^{\alpha}(t)=e^{\frac{i}{\hbar} H_{\rm B} t } Y^{\alpha} e^{ - \frac{i}{\hbar} H_{\rm B} t }$.
Here ${\cal T}$ is the time-ordering operator.
$\langle \cdots \rangle $ denotes the average over the canonical
state of the bath at the inverse temperature $\beta$.
We assume that there is no correlation between $Y^{\alpha}$ and $Y^{\gamma}$
when $\alpha \neq \gamma$, i.e., $\braket{Y^{\alpha}(t) Y^{\gamma}}_{\beta} =0$.
The canonical correlation function of the isolated bath, $\braket{Y^{\alpha}(t) Y^{\alpha}}_{\beta}$,
is characterized by a decay time, $\tau_{\rm bath}$, which has to be shorter than the
timescale of the system relaxation to the asymptotic state $(\tau_{\rm bath} \ll \tau_{\rm relax})$~\cite{hone2009statistical}.
This means that the master equation, Eq.~(\ref{eq:QMEop}), is valid under the condition of  a
weak, but finite system-bath coupling. It is noted here that $\tau_{\rm bath}$ is not identical to $\omega_c^{-1}$.

An intuitively good choice of basis for the density matrix of the system is the Floquet basis~\cite{sambe1973steady},
\begin{equation}
\rho (t)=\sum_{i,j} \rho_{ij}(t) \ket{u_i(t)} \bra{u_j(t)}.
\end{equation}
The Floquet states $\ket{u_i (0)}$ and quasi-energies $\epsilon_i$ are defined as the eigensystem of the one-period propagator $U(T, 0)$,
\begin{equation}
\label{eq:quasi}
U(T, 0) \ket{u_i (0)}=e^{-\frac{i}{\hbar} \epsilon_i T} \ket{u_i (0)},
\end{equation}
where $-\hbar \Omega/2 \leq \epsilon_i < \hbar \Omega/2$, $\Omega=2\pi/T$.
The Floquet state at any time $t$ can then be obtained by propagating the initial state $\ket{u_i(0)}$ with the propagator $U(t,0)$ and the appropriate phase factor~\cite{grifoni1998driven},
\begin{equation}
\ket{u_i (t)}= e^{\frac{i}{\hbar} \epsilon_i t } U(t, 0)\ket{u_i (0)}.
\end{equation}

Rewritten in the Floquet basis, the quantum master equation (\ref{eq:QMEop}) reads~\cite{grifoni1998driven},
\begin{eqnarray}
\frac{d\rho_{ij}(t)}{dt}=& -\frac{i}{\hbar} (\epsilon_i -\epsilon_j ) \rho_{ij}(t)-\frac{\lambda^2}{\hbar^2} \sum_{\alpha} \sum_{l,m} \sum_{s=-\infty}^{\infty}\sum_{s'=-\infty}^{\infty} e^{i(s+s')\Omega t} \nonumber\\
&\times \{ G^{\alpha}(\omega_{lm}^s) X^{\alpha}_{lms} X^{\alpha}_{ils'} \rho_{mj}(t) -G^{\alpha}(\omega_{im}^{s})X^{\alpha}_{ims} X^{\alpha}_{ljs'} \rho_{ml}(t) \nonumber\\
&-G^{\alpha} (-\omega_{lj}^s)^* X^{\alpha}_{ljs} X^{\alpha}_{ims'} \rho_{ml}(t) +G^{\alpha}(-\omega_{lm}^s)^* X^{\alpha}_{lms} X^{\alpha}_{mjs'} \rho_{il}(t) \},\nonumber \\
\label{eq:qmefl}
\end{eqnarray}
where frequencies $\omega_{lm}^s=(\epsilon_l -\epsilon_m )/\hbar + s \Omega$, and the
Fourier components of bath correlation functions are defined as
\begin{equation}
G^{\alpha}(\omega) =\int_0^{\infty} \braket{Y^{\alpha}(\tau) Y^{\alpha}}_{\beta} e^{-i \omega \tau} d\tau
\label{eq:corr}
\end{equation}
and Fourier coefficients of the matrix elements of the operator $X^{\alpha}$ read
\begin{equation}
X^{\alpha}_{lms}=\frac{1}{T} \int_0^T \bra{u_l (t)} X^{\alpha} \ket{u_m (t)} e^{-i s \Omega t} dt .
\end{equation}

Because of the linearity of the master equation and time-periodicity of its right-hand side, its asymptotic
solution, $\rho^{A}(t) = \lim_{N \rightarrow \infty}\rho(t + NT)$, becomes explicitly time-periodic,
$\rho^{A}(t + T) = \rho^{A}(t)$.
Below we address the asymptotic solution only (we are not considering the transients)
and hence the label `$A$' is  dropped from hereon.

The asymptotic state can be expressed by using the Fourier expansion of the elements of the density matrix $\rho(t)$, reading
\begin{equation}
\rho_{ij}(t)\equiv \rho_{ij}^{A}(t)=\sum_{s=-\infty}^{\infty} \rho_{ij}^s e^{is\Omega t}.
\end{equation}
Upon substituting this expansion into Eq.~(\ref{eq:qmefl}) we find the identity:
\begin{eqnarray}
0=&i \omega_{ij}^s\rho_{ij}^s+\frac{\lambda^2}{\hbar^2} \sum_\alpha \sum_{l,m} \sum_{s'=-\infty}^{\infty}\sum_{s''=-\infty}^{\infty} \Big\{ G^{\alpha} (\omega_{lm}^{s'}) X^{\alpha}_{lms'} X^{\alpha}_{il[s-(s'+s'')]} \rho_{mj}^{s''}\nonumber\\
&-G^{\alpha}(\omega_{im}^s) X^{\alpha}_{ims'} X^{\alpha}_{lj[s-(s'+s'')]} \rho_{ml}^{s''} -G^{\alpha} (-\omega_{lj}^s)^* X^{\alpha}_{ljs'} X^{\alpha}_{im[s-(s'+s'')]} \rho_{ml}^{s''}\nonumber\\
&+G^{\alpha} (-\omega_{lm}^s)^* X^{\alpha}_{lms'} X^{\alpha}_{mj[s-(s'+s'')]} \rho_{il}^{s''} \Big\}. \label{equations}
\end{eqnarray}
The second term on the r.h.s. of Eq.~(\ref{equations}) is of the order
$\lambda^2$, so in order to fulfill the equality,
the element $\rho_{ij}^s$ has to be small (by absolute value) when $\omega_{ij}^s$ is large~\cite{hone2009statistical}.
Thus, we use the approximation that for the set $\omega_{ij}^s\neq 0$,
\begin{equation}
\forall |\omega_{ij}^s| > \omega_{\rm trunc},   \quad \rho_{ij}^s=0, \label{approximation}
\end{equation}
where $\omega_{\rm trunc}$ is a truncation frequency. Put differently,
this approximation means that we set to zero those Fourier components of
the density matrix element $\rho_{ij}(t)$
for which the absolute value of $\omega_{ij}^s$ is larger than $\omega_{\rm trunc}$.
The approximation in Eq.~(\ref{approximation}) allows us to go beyond the RWA (within which
all $\rho_{ij}^s$ are zero except when $\omega_{ij}^s=0$). The RWA
is justified in the limit of infinitesimal coupling $\lambda \rightarrow 0$
and, evidently, the corresponding asymptotic state does not depend on the
coupling strength $\lambda$~\cite{ketzmerick2010statistical, kohn2001periodic, hone2009statistical}.
In contrast, our scheme shows the dependence on $\lambda$ for the asymptotic state.
It also allows us to go beyond the so-called `moderate RWA'~\cite{grifoni1998driven},
where only the $s=0$ mode is kept~\cite{blumel1991dynamical,oelschlagel1993a,oelschlagel1993b,gasparinetti2013environment}.

\section{Effective Floquet-Gibbs states}\label{Floquet-Gibbs}
In this section we define the notion of effective Floquet-Gibbs states.
In Ref.~\cite{shirai2015condition} Floquet-Gibbs states were introduced within the RWA framework,
-- under the assumption that all three conditions (i - iii) [see Sec.~\ref{intro}] hold --, by using
the Floquet Hamiltonian $H_{\rm F}^{[t]}$,
\begin{equation}
e^{-\frac{i}{\hbar} H_{\rm F}^{[t]} T}:={\cal T} e^{-\frac{i}{\hbar} \int_t^{t+T} H_{\rm S}(\tau) d\tau}.
\label{eq:floquet}
\end{equation}
The corresponding Floquet-Gibbs state is given by
\begin{equation}
\rho(t) =\frac{e^{-\beta H_{\rm F}^{[t]}}}{{\rm Tr}\left[e^{-\beta H_{\rm F}^{[t]}}\right]}.
\label{density_st}
\end{equation}
The Floquet Hamiltonian (\ref{eq:floquet}) is however not suitable when condition (i) is violated, as discussed in the introduction.

Here, we assume that the system is subjected to a fast and strong driving field,
\begin{equation}
H_{\rm S}(t)=H_0 +\xi H_{\rm ex}(t), \quad \int_0^T H_{\rm ex}(t) dt=0,
\end{equation}
where $\hbar \Omega$ and $\xi$ are much larger than a single site energy, e.g., the energy for flipping the spin at one site of the spin chain model,
but is smaller than the width of the energy spectrum of $H_0$.
This leads to the violation of condition (i).
Throughout this paper, we assume that condition (ii), i.e., $[H_{\rm ex}(t_1), H_{\rm ex}(t_2)]=0$, is satisfied [see Eq.~(\ref{eq:spin3})]. Hence it is convenient to transform to a rotating frame.
The states in the rotating frame, $|\psi^R(t ) \rangle$,
are related to the states in the original frame, $|\psi (t )\rangle$, through a unitary transformation
\begin{equation}
|\psi (t) \rangle = {\cal T}e^{-i \frac{\xi}{\hbar} \int_0^t H_{\rm ex}(\tau) d\tau}|\psi^R (t ) \rangle =V (t) |\psi^R (t )\rangle. \label{rotating_frame}
\end{equation}
Thus, the system Hamiltonian $H_{\rm S}(t)$ in the rotating frame is transformed to
\begin{equation}
H_{\rm S}^R (t)=V^{\dagger}(t) \left( H_{\rm S}(t) -i \hbar \frac{\partial}{\partial t}\right) V(t)=V^{\dagger}(t) H_0 V(t).
\end{equation}
It is noted here that the condition (ii) ensures the time periodicity of $V(t)$, and hence $H_{\rm S}^R (t)$ is time periodic.

While the amplitude of the driving field in $H_{\rm S}(t)$ is given by $\xi$, which is very large,
the amplitude of the oscillating term in $H_{\rm S}^R (t)$ is not strong.
To illustrate this point, we present a driven single spin-1/2 system as an example, whose Hamiltonian is given by $H_{\rm S}(t)=h_z S^z +\xi \cos \Omega t S^x$.
The Hamiltonian in the rotating frame reads
\begin{equation}
H_{\rm S}^R(t)=h_z \left[ S^z \cos \left( \frac{\xi}{\hbar\Omega} \sin \Omega t \right) +S^y \sin \left( \frac{\xi}{\hbar\Omega} \sin \Omega t \right) \right].
\end{equation}
The oscillating strength of the rotated Hamiltonian is bounded by $h_z$ (and not $\xi$ as in the original frame).
Thus, since $h_z \ll \xi$, one could expect that in the rotating frame the Magnus expansion,
which is a high frequency expansion of the Floquet Hamiltonian~\cite{blanes2009magnus},
provides an accurate description with fewer terms in
the expansion as compared to the expansion in the original frame.

In order to uniquely define the effective energies we introduce an $n$-th order effective (truncated) Floquet Hamiltonian,
\begin{equation}
H_{\rm F}^{[t](n)} = \frac{1}{T}\sum_{k=1}^{n} \Omega_{k}^{[t]}(T),
\label{eq:magnus}
\end{equation}
with the summands defined by
\begin{eqnarray}
\label{Magnus1}
\Omega_{1}^{[t]}(T)=&\int_{t}^{t+ T } H_{\rm S}^R (\tau' ) d\tau',\\
\label{Magnus2}
\Omega_{k}^{[t]}(T)=&\sum_{j=1}^{k-1} \frac{(-i)^j B_j}{\hbar^j j!}\sum_{s_1=1}^{\infty} \cdots \sum_{s_j=1}^{\infty} \delta_{\sum_{i=1}^j s_i, k-1} \nonumber\\
&\times \int_{t}^{t+ T} [ \Omega_{s_1}^{[t]}(\tau' ) \cdots [\Omega_{s_j}^{[t]}(\tau' ), H_{\rm S}^R (\tau' )] \cdots ] d\tau',~~~~~\forall k\geq 2,
\end{eqnarray}
where $B_j$ is the $j$-th Bernoulli number and $\delta$ is
the Kronecker delta~\cite{blanes2009magnus}.
The term $\Omega_{k}^{[t]}(T)$ is of the order of $T^k$.
First of all, the ambiguity in definition of effective energies is gone: energies of the effective Floquet Hamiltonian, Eq.~(\ref{eq:magnus}),
are uniquely defined.
In addition, this step provides an efficient basis to express the asymptotic density matrix.
It is noteworthy that Magnus expansion is continuously bridging two extremes.
When used in the complete form, it yields the basis of Floquet states. The truncation right
after the first term corresponds to the diabatic basis~\cite{hone2009statistical}.

When a periodically modulated system is coupled to a dissipative environment, the interaction with the environment suppresses the heating to infinite temperature, which usually takes place in isolated systems~\cite{lazarides2014equilibrium, dalessio2014long-time, ponte2015periodically}. Under the high-frequency driving the heating rate is low. Thus, when the coupling is weak but finite, the dissipation rate may become larger than the heating rate. This will cause the system to equilibrate and not heat to infinite temperature. In this situation the eigenbasis of the effective (truncated) Floquet Hamiltonian is preferable over the Floquet basis. It is therefore reasonable to probe the idea of the asymptotic density matrix having the Gibbs
form in the basis of the \textit{effective} Floquet Hamiltonian, with some -- not yet known -- value of truncation $n:=n_{\rm{eff}}$.
The so defined effective Floquet-Gibbs density matrix has the form:
\begin{equation}
\rho_{\rm EFG}(t) =\frac{e^{-\beta H_{\rm F}^{[t](n_{\rm{eff}})}}}{{\rm Tr}\left[ e^{-\beta H_{\rm F}^{[t](n_{\rm{eff}})}}\right]},\label{density}
\end{equation}
where $\beta$ is the inverse temperature of the heat bath.

The sufficient condition for the convergence of the Magnus expansion is $2 \| H_{0} \| < \hbar\Omega$~\cite{blanes2009magnus}.
It is not met in our case and hence the Magnus expansion does not converge in general.
Recently, a definition for the optimal value $n_{\rm{eff}}$ was proposed in Ref.~\cite{kuwahara2016floquet}.
This value minimizes the deviation of the eigenspectrum of the Floquet Hamiltonian from that of
its truncated version,
\begin{eqnarray}
\Delta= \max_{\ket{\psi}\atop (\bra{\psi} \psi \rangle =1)} \left\| \left( e^{-\frac{i}{\hbar} H_{\rm F}^{[0]} T} -e^{-\frac{i}{\hbar} H_{\rm F}^{[0](n)} T} \right) \ket{\psi} \right\|.
\label{eq:delta}
\end{eqnarray}
This deviation can be evaluated by diagonalizing the operator in parenthesis
and then taking the maximal eigenvalue from the eigenspectrum.

It has been shown~\cite{kuwahara2016floquet} that the smallness of $\Delta$ ensures
a good description of the long-time transient dynamics by an effective Floquet Hamiltonian for any initial state.
We adapted this idea and use $n_{\rm{eff}}$ to construct the effective Floquet Hamiltonian. Its eigenvectors
should be used then to obtain the effective Floquet-Gibbs state.
When condition (i) is satisfied, e.g., in the limit of fast modulations,
$n_{\rm eff} \rightarrow \infty$ and $H_{\rm F}^{[t] (n_{\rm eff})}=H_{\rm F}^{[t]}$, the standard Floquet-Gibbs state, Eq.~(\ref{density_st}), is recovered, since the Magnus expansion is convergent in this case.
\begin{figure*}[t]
\begin{center}
\includegraphics[width=0.5\textwidth]{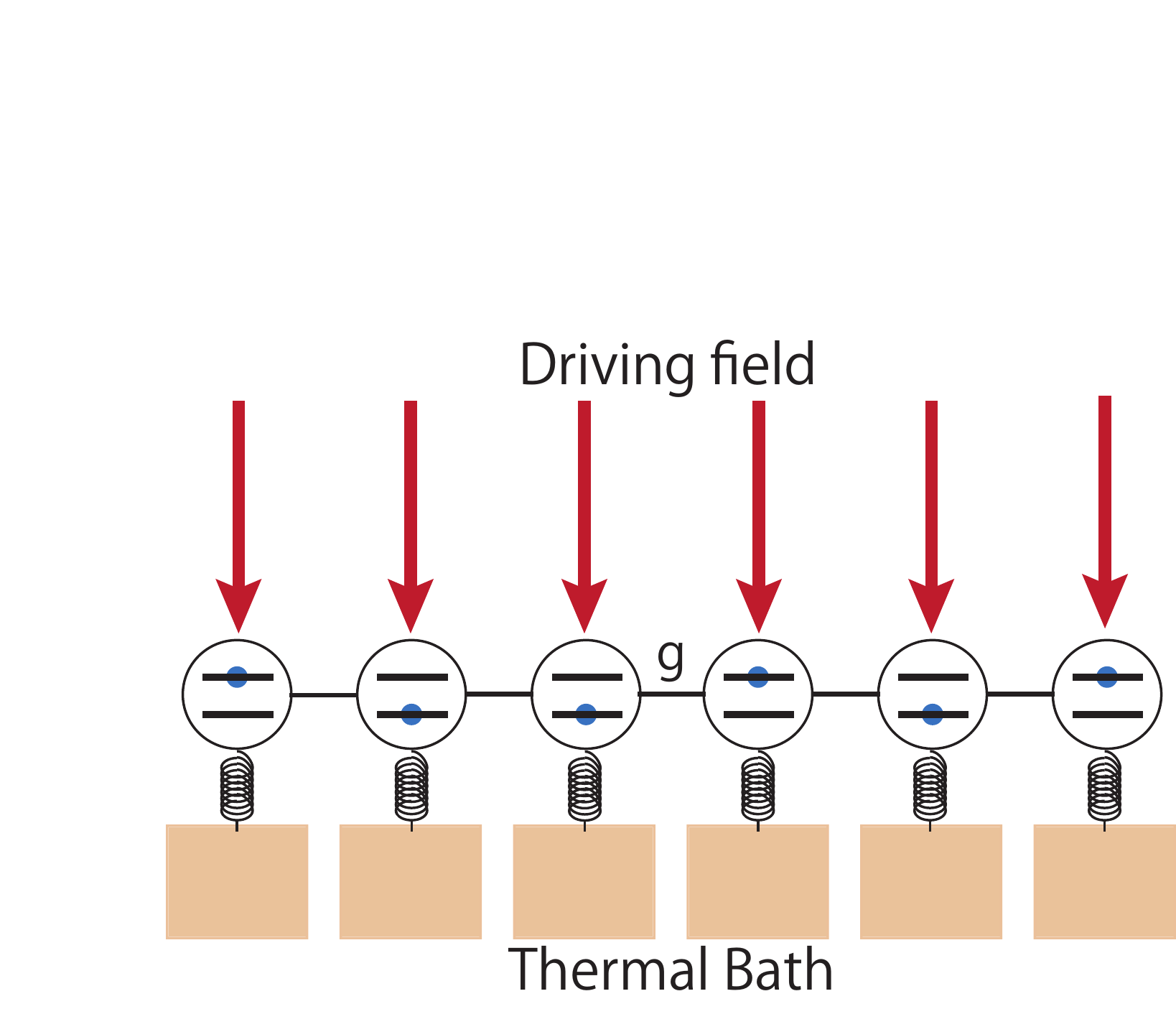}
\end{center}
\caption{(Color online) A spin chain subjected to a time-periodic magnetic field being coupled to an environment.
Each spin site is connected to an independent heat bath.
The baths have the same properties and are all kept at the same inverse temperature $\beta$.}
\label{schematic}
\end{figure*}

So far we have introduced two truncation parameters, $\omega_{\rm trunc}$ and $n_{\rm eff}$.
The cutoff frequency $\omega_{\rm trunc}$ is necessary to deal with the situation with a large system Hilbert space.
The second truncation parameter $n_{\rm eff}$, for the Magnus expansion,
is needed to obtain an effective Hamiltonian and hence to construct an effective Floquet-Gibbs state.

In order to estimate how close the reduced density matrix $\rho^R (t)$ obtained in the rotating frame,
see Eqs.~(\ref{equations}) and~(\ref{approximation}), is to the effective Floquet-Gibbs state, Eq.~(\ref{density}),
we employ the trace distance
\begin{equation}
\Delta {\rm Prob}^{[t]}={\rm Tr} | \rho^R (t) - \rho_{\rm EFG}(t) |.~\label{difference}
\end{equation}
The trace distance provides an upper bound to the accuracy of expectation values of a system observable $\hat{O}$,
\begin{equation}
|{\rm Tr}(\hat{O} \rho^R (t))-{\rm Tr}(\hat{O} \rho_{\rm EFG} (t))| \leq \Delta {\rm Prob}^{[t]} \| \hat{O} \|,
\end{equation}
where $\| \hat{O} \|$ is the operator norm of $\hat{O}$.

\section{Conditions for the realization of the effective Floquet-Gibbs states}
\label{condition}

As argued in the introduction the Floquet-Gibbs state can be realized even when condition (i) and/or (iii) become broken. In this section we elaborate further on the conjecture for the \emph{alternative} conditions when conditions (i) and (iii) are violated. The breaking of these conditions requires a careful consideration of the various timescales governing the system of interest, driving field, and the heat bath. When condition (i) is broken, the system may heat up due to the resonance with the driving field,
and hence the timescale for heating $\tau_{\rm heat}$ becomes finite.
However, under the high-frequency driving field the heating process simultaneously excites multiple sites and acquires  extremely long times to occur~\cite{bukov2015universal, kuwahara2016floquet, mori2016rigorous, abanin2015asymptotic, abanin2015effective}. Thus, if $\tau_{\rm heat} \gg \tau_{\rm relax}$, it is expected that the resonant heating is suppressed by the dissipation due to the heat bath so that a violation of condition (i) will not affect the realization of the effective Floquet-Gibbs state.

The other timescales are the period of the driving field $T$ and the inverse of the cutoff frequency $\omega_c^{-1}$. In order to understand the connection between these timescales and condition (ii) we transform the total Hamiltonian into the rotating frame,
by using the unitary transformation, Eq.~(\ref{rotating_frame}),
\begin{equation}
H^{R}(t)=H_{\rm S}^R (t)+H_{\rm B}+\lambda H_{\rm I}^{R}(t),
\end{equation}
where $H_{\rm I}^{R}(t)$ is given by
\begin{equation}
H_{\rm I}^{R}(t)=V^{\dagger}(t) H_{\rm I} V (t).
\end{equation}
The periodic modulations of the interaction Hamiltonian $H_{\rm I}^R (t)$
excites the fast harmonic modes of the heat bath,
whose frequencies are close to multiples of the driving frequency.
Hence, the response of the system to this field-enforced bath dynamics can affect the asymptotic state and induce deviations from the  effective Floquet-Gibbs form.
However, if $\omega_c \ll \Omega$, it is expected that the excitations induced in the heat bath due to the driving are weak, such that
it does not largely affect the asymptotic state of the system.
Alternatively, when the condition (iii) is imposed,
$H_{\rm I}^R (t)$ becomes time-independent in the rotating frame
and thus it cannot excite the resonant modes in the bath.

Overall, the restrictive conditions (i) and (iii), due to the above reasonings, are replaced by conditions which do not severely limit the system or the driving field. These new conditions for the emergence of the \emph{effective} Floquet-Gibbs state can be summarized as,
\begin{itemize}
\item[] (1) $\tau_{\rm heat} \gg \tau_{\rm relax}$;
\item[] (2) $[H_{\rm ex}(t_1), H_{\rm ex}(t_2)]=0$ for any $t_1$ and $t_2$;
\item[] (3) $\omega_c \ll \Omega$ or $[H_{\rm I}, H_{\rm ex}(t)]=0$.
\end{itemize}
We emphasize that condition (ii) $\equiv$ (2) remains as it is due to its less restrictive nature.

\section{Spin chain model}
\label{spinchain}
\begin{figure}[t]
\begin{center}
\includegraphics[width=0.5\textwidth]{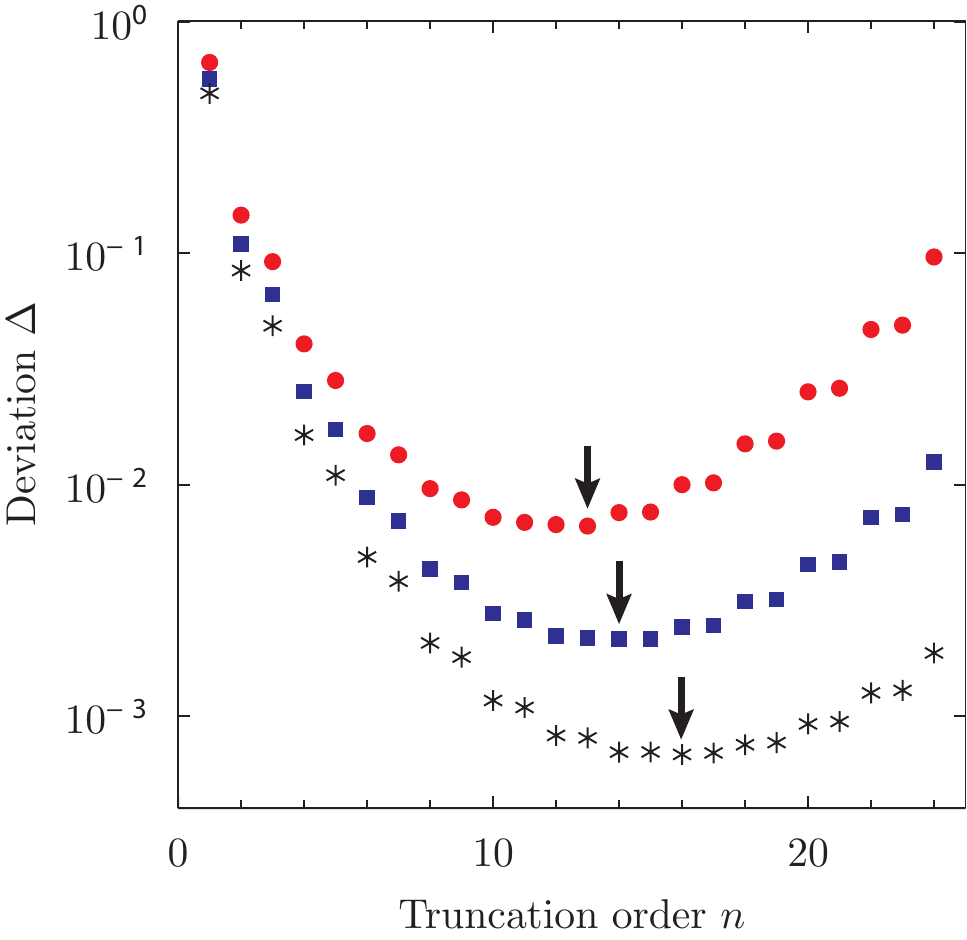}
\end{center}
\caption{(Color online) Deviation $\Delta$, Eq.~(\ref{eq:delta}), vs the truncation order
$n$, for the spin chain model, Eqs.~(\ref{eq:spin1}-\ref{eq:spin3}).
Three values of the driving frequency, $\hbar\Omega=4.2h_z$ ({\color{red}$\bullet$}),
$4.6h_z$ ({\color{blue}$\subHalmos$}), and $5.0h_z$ ($\ast$) were used in simulations,
while keeping the ratio $\xi/\hbar\Omega=2/3$ fixed. Arrows indicate the
effective truncation order $n_{\rm{eff}} = 13$ ($\hbar\Omega = 4.2h_z$), $14$ ($\hbar\Omega = 4.6h_z$) and $16$ ($\hbar\Omega = 5.0h_z$).}
\label{sx}
\end{figure}

In order to probe the idea of effective Floquet-Gibbs state,
we use as a testbed a quantum spin$-1/2$ chain being subjected to a
strong high-frequency driving, see the sketch in Fig.~\ref{schematic},
\begin{eqnarray}
\label{eq:spin1}
H_{\rm S}(t) = H_0 + \xi H_{\rm ex} (t), \\
\label{eq:spin2}
H_0= h_z \sum_{i=1}^N S_i^z -g \sum_{i=1}^{N-1} S_i^x S_{i+1}^x+h_x \sum_{i=1}^N S_i^x, \\
\label{eq:spin3}
H_{\rm ex}(t)= \cos\left( \Omega t\right) \sum_{i=1}^N S_i^x.
\end{eqnarray}
Here $h_z$ and $h_x$ are constant components of the magnetic field acting along $z$ and $x$ directions, respectively, and
$g$ is the coupling between neighboring spins.
In addition, there is a time-periodic component of the field
acting along $x$ direction, introduced through the term $H_{\rm ex}(t)$.
This model describes a quasi-one-dimensional
Ising ferromagnet~\cite{coldea2010quantum} or
a chain of interacting superconducting qubits~\cite{grajcar2006four}.
Throughout this work we use the following set of
parameters: $N=6$, $g=0.75h_z$, $h_x=0.7h_z$, and keep the $\xi/(\hbar\Omega) = 2/3$ fixed.
These set of parameters fix the width of the energy spectrum of $H_0$ to $\Delta_0 =7.6h_z$. This allows us to study
the regime when the frequency of the driving field $\Omega$ is smaller than $\Delta_0/\hbar$
and thus condition (i), see Sec.~\ref{intro}, is violated. We employ three different values for driving frequency, i.e., $\hbar\Omega=4.2h_z, 4.6h_z$, and $5.0h_z$. These values are all smaller than $\Delta_0/\hbar$, but larger than the characteristic frequency of a single spin.
We choose the truncation frequency, $\hbar \omega_{\rm trunc}$, to be equal to $10 h_z$,
and have verified that $\Delta{\rm Prob}^{[t]}$ has converged by varying the truncation frequency up to $15 h_z$ (see Fig.~\ref{fig1_supp} of \ref{trunc}).

Figure~\ref{sx} depicts $\Delta$ [Eq.~(\ref{eq:delta})] as a function of the truncation order $n$.
The deviation initially decreases with $n$ but then, after reaching a minimum, it increases again.
The minimum determines the value of $n_{\rm{eff}}$.
From hereon for notational simplicity we will suppress the explicit dependence on $n_{{\rm eff}}$ and implicitly assume the values $n_{{\rm eff}} = 13, 14,$ and $16$
for driving frequencies $\hbar\Omega = 4.2h_z, 4.6h_z,$ and $5.0h_z$, respectively.

\begin{figure}[t]
\begin{center}
\includegraphics[width=0.5\textwidth]{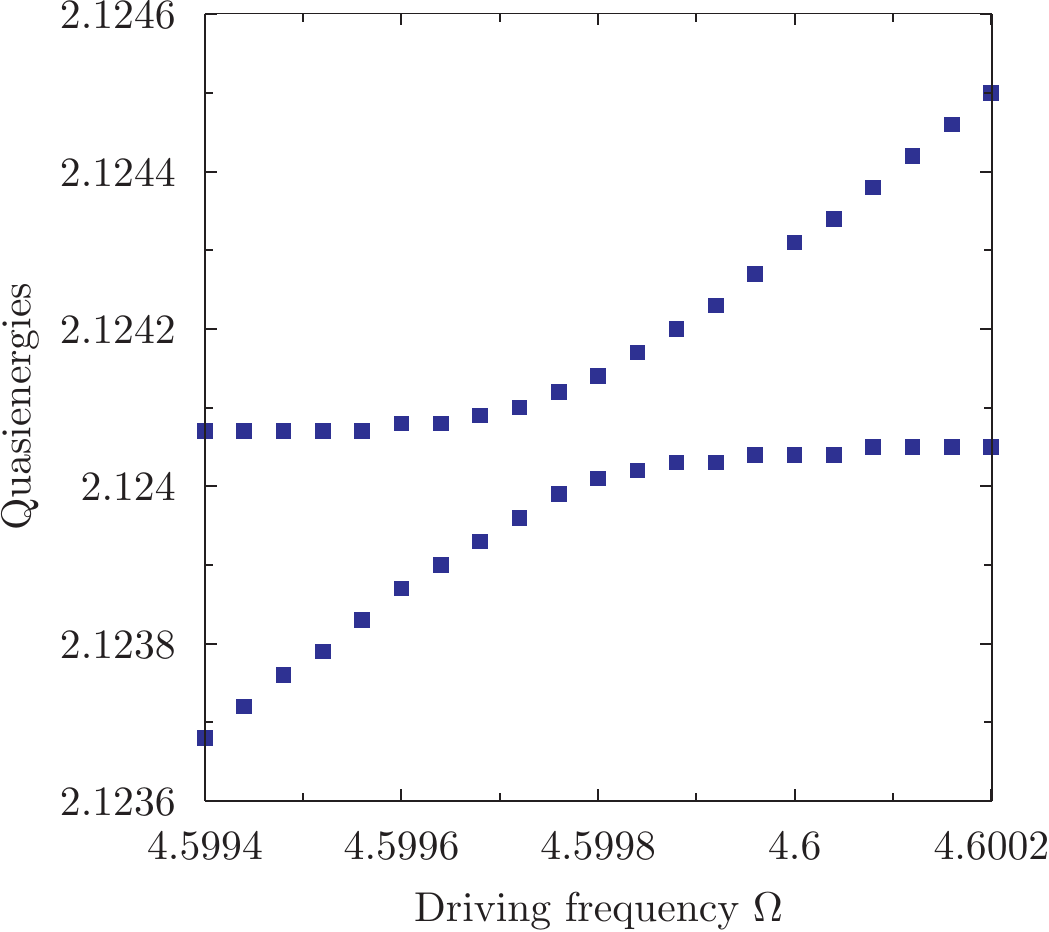}
\end{center}
\caption{(Color online) The pair of quasi-energies (blue squares) that lead to the resonance effect vs driving frequency $\hbar \Omega/h_z$. The quasi-energies show an avoided crossing around $\hbar \Omega=4.6 h_z$.}
\label{avoided}
\end{figure}

At $\hbar\Omega = 4.6h_z$ we observe the resonance effect (see the next section), which appears as an avoided crossing for a pair of quasi-energies. In Fig.~\ref{avoided} we plot the pair of the quasi-energies in resonance. The inverse of the quasi-energy gap at the avoided crossing represents the heating timescale $\tau_{\rm heat}$ \cite{hone2009statistical}.

The quantum master equation, introduced in Sec.~\ref{redfield}, requires information about the nature of the bath in terms of the correlation function described in Eq.~(\ref{eq:corr}). To specify the function, we follow the standard prescription and consider a heat bath
consisting of a set of independent harmonic oscillators ~\cite{nakajima1958quantum, zwanzig1960ensemble}. The heat bath is described by a Hamiltonian
\begin{equation}
H_{\rm B}= \sum_{i=1}^{N} \sum_{\alpha} \left( \frac{p_{i}^{\alpha 2}}{2 m_{\alpha}} +\frac{m_{\alpha} \omega_{\alpha}^2}{2} x_{i}^{\alpha 2} \right),
\end{equation}
where $x^{\alpha}_{i}$ and $p^{\alpha}_{i}$ are canonical variables of the $\alpha$th oscillator in the $i$th heat bath,
and $m_\alpha$ and $\omega_{\alpha}$ are the mass and frequency of the oscillator, respectively.
The sum over spin site $i$ indicates that each of the spins is connected to an individual heat bath which is non-interacting and uncorrelated with any other bath~\cite{fischer2013coherence}.

By use of Eq.~(\ref{eq:Ham}) we set the system-bath interaction Hamiltonian $H_{\rm I}$,
\begin{equation}
H_{\rm I}=\sum_{i=1}^{N} \sum_{\alpha} c_{\alpha} x_i^{\alpha} \otimes (a^x S_i^x +a^y S_i^y +a^z S_i^z ),
\label{eq:HI}
\end{equation}
where the scalar set $(a^x, a^y, a^z)$ determines whether condition (iii) [refer Sec.~\ref{intro}], $[H_{\rm I}, H_{\rm ex}(t)]=0$, is satisfied or not.
The baths are characterized by a spectral density that we choose to be of the Ohmic form,
\begin{equation}
\label{eq:cutoff}
J (\omega) =\sum_{\alpha}\frac{\pi c_{\alpha}^{2}}{2 m_{\alpha} \omega_{\alpha}} \delta (\omega -\omega_{\alpha}) =\tilde{\gamma}\omega e^{-\frac{\omega}{\omega_c}}.~\label{spectral}
\end{equation}

Throughout this work we have explicitly used the parameter $\lambda$ as a dimensionless parameter to indicate the strength of the system-bath coupling. The actual dissipation strength is $\gamma = \lambda^{2}\tilde{\gamma}$.
Thus the timescale for relaxation is given by $\tau_{\rm relax}=\gamma^{-1}$, which is controlled by $\lambda^2$.
For simplicity from hereon we will only vary the parameter $\lambda^{2}$ and set $\hbar\tilde{\gamma} = h_z$. In order to probe the dependence of timescales that lead to the emergence of the effective Floquet-Gibbs state in sec.~\ref{finitecoupling} we present the dependence of $\Delta {\rm Prob}^{[t]}$ on the system-bath coupling strength $\lambda^2$. Here we show that the effective Floquet-Gibbs state of the spin chain model appears in the regime $\tau_{\rm heat} \gg \tau_{\rm relax}$ [condition (1) of Sec.~\ref{condition}]. Section~\ref{timescale} shows the dependence of $\Delta {\rm Prob}^{[t]}$ on the cutoff frequency $\omega_c$ and in this case the effective Floquet-Gibbs state emerges in the regime $\omega_c \ll \Omega$ [condition (3) of Sec.~\ref{condition}].

\section{Asymptotic states: Dependence on dissipation strength}
\label{finitecoupling}
\begin{figure}[t]
\begin{center}
\includegraphics[width=0.5\textwidth]{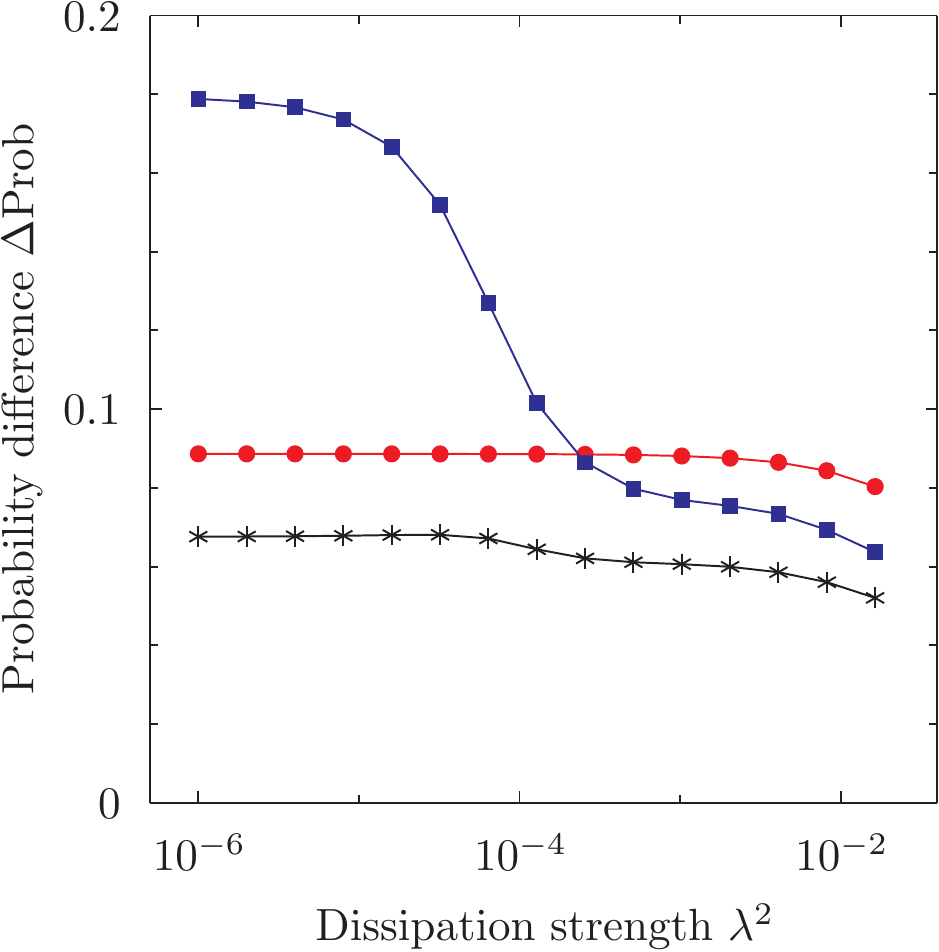}
\end{center}
\caption{(Color online) The dependence $\Delta{\rm Prob}\equiv \Delta{\rm Prob}^{[0]}$, Eq.~(\ref{difference}),
on the system-bath coupling (squared) $\lambda^2$ (dissipation strength) for
three values of the driving frequency, $\hbar\Omega=4.2h_z$ ({\color{red}$\bullet$}),
$4.6h_z$ ({\color{blue}$\subHalmos$}), and $5.0h_z$ ($\ast$).
There is a resonance (see text) for $\hbar\Omega=4.6h_z$, which is responsible for a strong deviation
from the effective Floquet-Gibbs state in the weak-coupling limit.
The deviation decreases upon the increase of the coupling strength.
The cutoff frequency, Eq.~(\ref{eq:cutoff}), $\hbar\omega_c = 100h_z$.
Other parameters are same as in Fig.~\ref{sx}.}
\label{DIFGa}
\end{figure}
In this section we discuss the asymptotic solutions when condition (i) [refer Sec.~\ref{intro}] is violated. We set the system-bath interaction Hamiltonian $H_{\rm I}$, Eq.~(\ref{eq:HI}) with $a^{x}=1$ and $a^{y}=a^{z}=0$, to a form that commutes with $H_{\rm ex}(t)$, i.e.,
\begin{equation}
H_{\rm I}=\sum_{i=1}^{N} \sum_{\alpha}c^{\alpha} x_i^{\alpha} \otimes S_i^x.
\label{eq:HIcom}
\end{equation}
This choice satisfies
condition (iii), see Sec.~\ref{intro}, and hence allows us to focus
on the effects of finite dissipation of strength $\lambda^2$, cf. Eq.(1), especially when condition (i) is violated. We fix the cutoff frequency $\hbar\omega_c=100h_z$. As far as only condition (i) is concerned,
the value of the cutoff frequency does not play a crucial role (it is, however, related to condition (iii), which we address in the next section). The inverse temperature of the heat baths is set at $\beta=1h_z^{-1}$.

\begin{figure}[t]
\begin{center}
\includegraphics[width=0.5\textwidth]{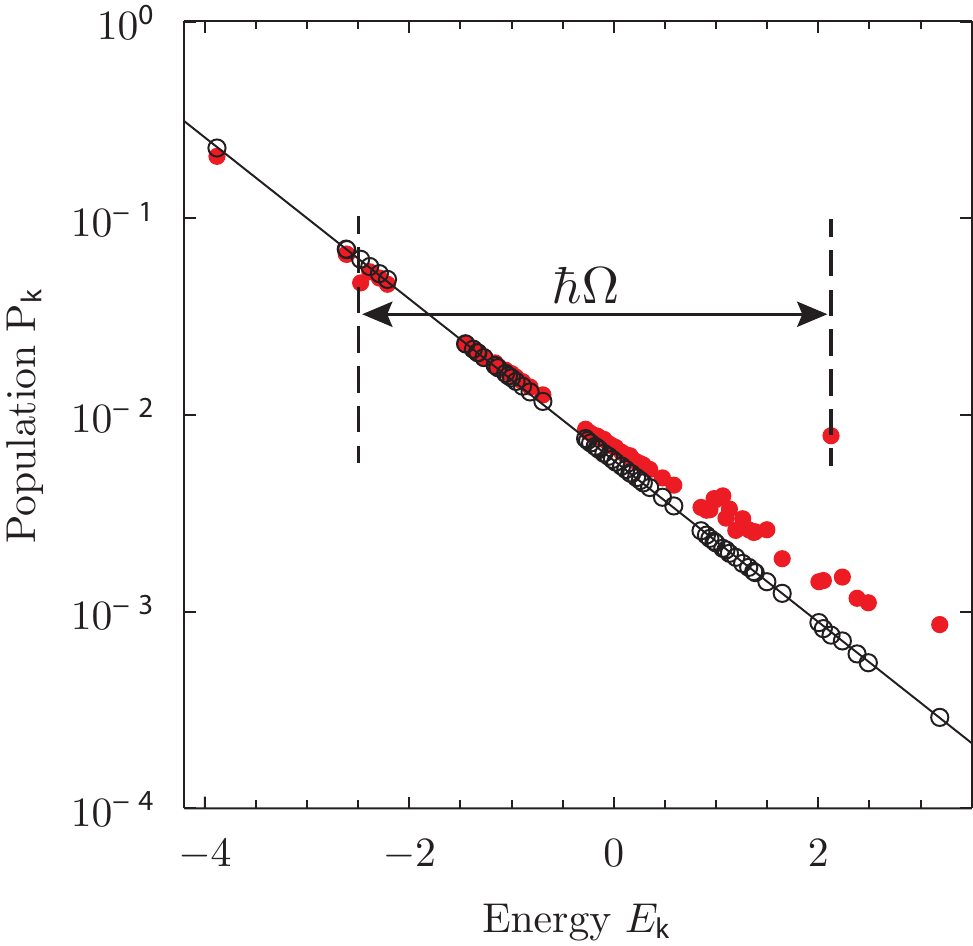}
\end{center}
\caption{(Color online)The populations ${\rm P}_k$ of the eigenstates
of the effective Floquet Hamiltonian of the
driving frequency $\hbar\Omega=4.6h_z$ and two values of the coupling,
$\lambda^{2} = 10^{-6}$ ({\color{red}$\bullet$}) and $\lambda^2=10^{-2}$ ($\circ$).
In case of $\lambda^{2}=10^{-6}$ there are two
eigenstates resonantly coupled by the driving field. Their populations
deviate strongly from the Boltzmann distribution. Solid black line
corresponds to the Boltzmann distribution with $\beta_{\rm eff}=0.946h_z^{-1} < \beta = 1h_z^{-1}$.
Other parameters are same as in Fig.~\ref{DIFGa}.}
\label{DIFGb}
\end{figure}

Figure~\ref{DIFGa} depicts the dependence of the trace distance $\Delta {\rm Prob}$, Eq.~(\ref{difference}),  as a function of $\lambda^2$. As shown in Fig.~\ref{fig2_supp} of \ref{timedependence}, the trace distance is independent of $t$ and hence we suppress the super-script $[t]$ and plot $\Delta {\rm Prob} \equiv \Delta {\rm Prob}^{[0]}$ as a representative.

The values of $\Delta {\rm Prob}$ for $\lambda^2 = 10^{-6}$ are nearly identical to those obtained within the RWA.
We find that for the driving frequency $\hbar\Omega = 4.6h_z$, $\Delta {\rm Prob}$
reduces as the system-bath coupling increases.
This indicates that finite dissipation can push the asymptotic state closer to the effective Floquet-Gibbs
state. The large deviation in the weak-coupling limit originates from the resonance effect, i.e., when the
energy gap between two eigenvalues of the effective Floquet Hamiltonian is in resonance with the driving frequency.
The resonance is not observed for two other values, $\hbar\Omega=4.2h_z$ and $5.0h_z$, hence in these cases
the asymptotic density matrix can be reasonably approximated with Eq.~(\ref{density}).

In order to further probe into the asymptotic effective Floquet-Gibbs state, we calculate the populations,
\begin{eqnarray}
{\rm P}_k & = \bra{\phi_k} \rho^R (0) \ket{\phi_k},
\end{eqnarray}
where $\ket{\phi_k}$ is an eigenstate of the effective Floquet Hamiltonian with eigenvalue, $E_k$.
In Fig.~\ref{DIFGb}, we plot the distribution for $\hbar\Omega=4.6h_z$ (the resonance case) and two values of the coupling strength,
$\lambda^2 = 10^{-2}$ and $\lambda^2 = 10^{-6}$. For the weak coupling case we observe two energies of the effective Floquet Hamiltonian in resonance with the driving frequency $\Omega$.

For stronger coupling the resonance is completely suppressed and the distribution of populations is close to the Boltzmann distribution. This suppression of resonance effect has been previously observed in the diabatic basis~\cite{hone2009statistical}, wherein the authors found that the diabatic basis leads to a diagonalized form of the asymptotic state. This occurred only when the dissipation strength exceeds the quasi-energy splitting observed at the avoided crossing (see in Fig.~\ref{avoided}). For the strong coupling case in Fig.~\ref{DIFGb} we linearly fit the log dependence of the probability to yield the exponent $\beta_{\rm eff}$ which is smaller than the temperature of the bath, i.e., $\beta_{\rm eff}=0.946h_z^{-1} < \beta=1h_z^{-1}$. This observation is in agreement with the previous results obtained within the RWA~\cite{breuer2000quasistationary, ketzmerick2010statistical}. In these works the `effective temperature' $T_{\rm eff} = 1/k_{\mathrm{B}} \beta_{\rm eff} $ has also been found to
be higher than the actual temperature of the heat bath. The mechanism behind the `effective temperature', although very intriguing, is beyond the scope of this work.

\begin{figure}[t]
\begin{center}
\includegraphics[width=0.5\textwidth]{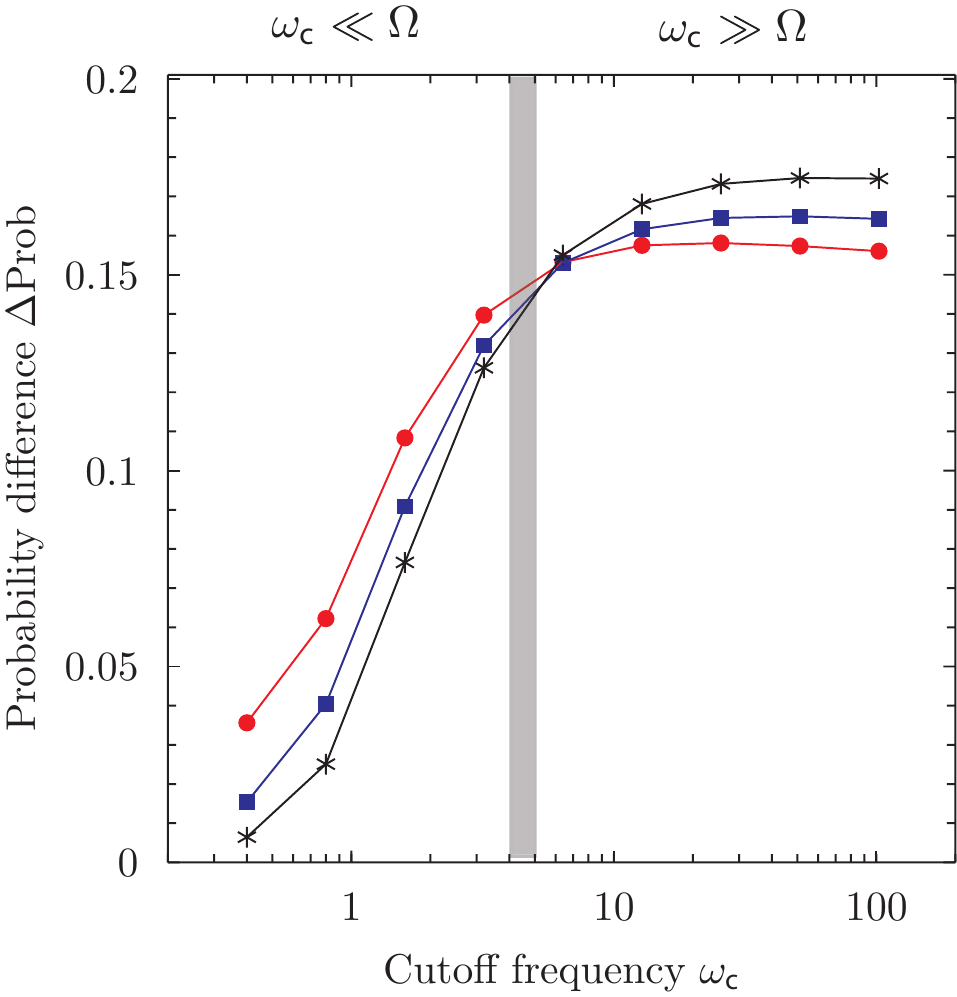}
\end{center}
\caption{(Color online) The difference $\Delta{\rm Prob}\equiv \Delta{\rm Prob}^{[0]}$, Eq.~(\ref{difference}),
as a function of the cutoff frequency $\omega_c$,
of the spectral density of the heat bath, Eq.~(\ref{eq:cutoff}), for three values of the driving
frequency, $\hbar\Omega=4.2h_z$ ({\color{red}$\bullet$}),
$4.6h_z$ ({\color{blue}$\subHalmos$}), and $5.0h_z$ ($\ast$) at  $\lambda^2=10^{-2}$.
Gray stripe marks the interval $\hbar\omega_c/h_z \in [4.2, 5]$.}
%The values of $n_{{\rm eff}}$ is same as that in Fig.~\ref{sx}.}
\label{w_c-dependence}
\end{figure}

\section{Asymptotic states: Dependence on the cutoff frequency of the heat bath}
\label{timescale}

Based on the conjecture in Sec.~\ref{condition}, in this section we analyze the
transition to the effective Floquet-Gibbs form under the variations of the cutoff frequency of the bath. Now we drop condition (iii), and address the general case where the interaction Hamiltonian, Eq.~(\ref{eq:HI}) with $a^{x}=a^{y}=1$ and $a^{z}=0$,
does not commute with the driving Hamiltonian, i.e. $[H_{\rm I}, H_{\rm ex}(t)] \neq 0$. Namely, we use the following interaction Hamiltonian:
\begin{equation}
H_{\rm I}=\sum_{i=1}^N \sum_{\alpha} c^{\alpha} x_{i}^{\alpha} \otimes (S_i^x +S_i^y).
\end{equation}
In this case too the inverse temperature of the heat baths is set to $\beta=1h_z^{-1}$. It is worth noting here that throughout this work we have neglected the counter-term that  appears in the Zwanzig-Caldeira-Leggett model~\cite{hanggi2005chaos,zwanzig1973nonlinear, caldeira1981influence, caldeira1983quantum}. In the present choice of the system-bath interactions, the counter-term plays no role since it is proportional to $S_i^{x2}$ or $(S_i^x+S_i^y)^2$ that cause a constant shift in the system potential.

Figure~\ref{w_c-dependence} presents the dependence of $\Delta {\rm Prob} \equiv \Delta{\rm Prob}^{[0]}$ on the cutoff frequency
for different values of the driving frequency at $\lambda^2=10^{-2}$. Similar to the last section we find that $\Delta{\rm Prob}^{[t]}$ is time-independent as shown in the Fig.~\ref{fig2_supp} of \ref{timedependence}.
When the cutoff frequency is large, $(\omega_c \approx 100h_z/\hbar \gg \Omega)$,
the deviation of the asymptotic state from the effective Floquet-Gibbs state is large, owing to the violation of the
condition (iii). On the other hand, for small cutoff frequency, $\omega_c \ll \Omega$, the asymptotic state is well described by the effective Floquet-Gibbs state, Eq.~(\ref{density}).

\begin{figure}[t]
\begin{center}
\includegraphics[width=0.5\textwidth]{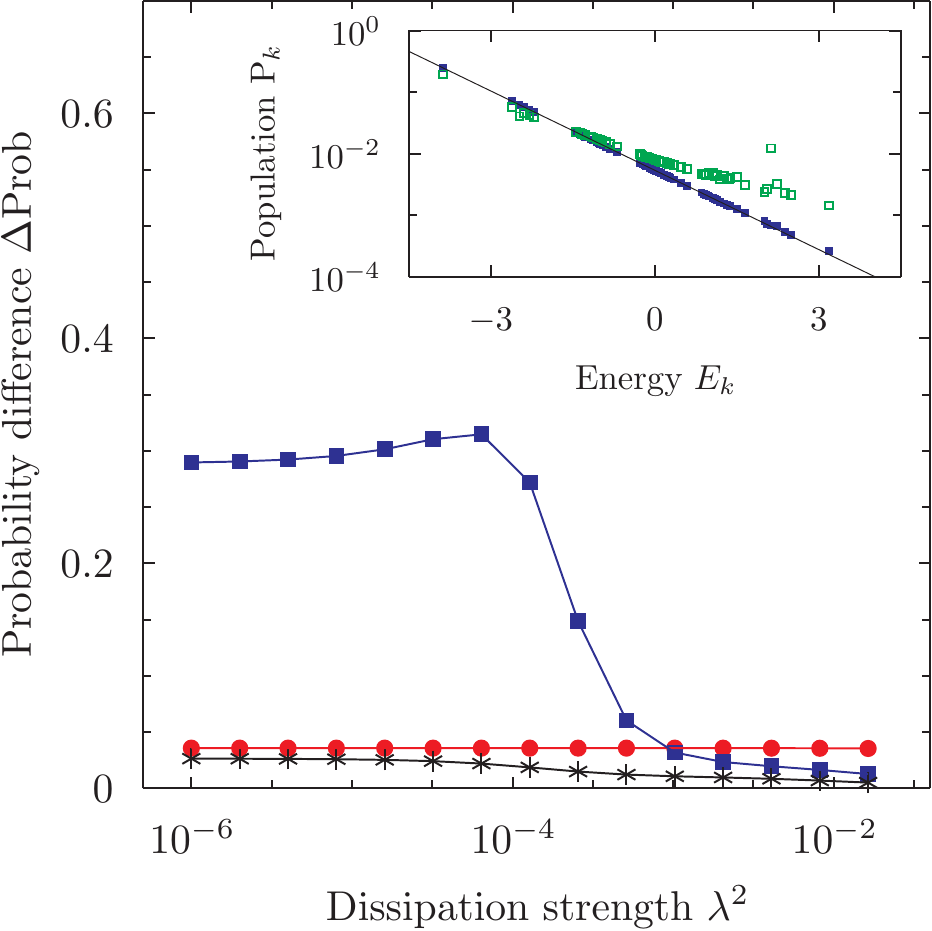}
\end{center}
\caption{(Color online)
The dependence of $\Delta{\rm Prob}\equiv \Delta{\rm Prob}^{[0]}$, Eq.~(\ref{difference}),
on the system-bath coupling $\lambda^2$
for $\hbar\Omega=4.2h_z$ ({\color{red}$\bullet$}),
$4.6h_z$ ({\color{blue}$\subHalmos$}), and $5.0h_z$ ($\ast$).
The cutoff frequency is $\hbar\omega_c = 0.4h_z$.
For $\hbar\Omega=4.6h_z$ there is a resonant coupling of two states
induced by the driving field, so that the dependence follows the scenario presented on Fig.~\ref{DIFGa}.
Inset: The populations ${\rm P}_k$ of eigenstates of the
effective Hamiltonian for $\hbar\Omega=4.6h_z$, for two values of dissipation
strength, $\lambda^2=10^{-2}$({\color{blue}$\subHalmos$}, blue closed boxes)
and $\lambda^2=10^{-6}$({\color{green}\protect\makebox[6pt]{$\square$}{\hspace{0.em}}}, green open boxes). The solid line is
the Boltzmann distribution with the exponent $\beta_{\rm eff} = 0.99$.
%The values of $n_{{\rm eff}}$ is same as that in Fig.~\ref{sx}.
}
\label{law-w_c}
\end{figure}

Figure~\ref{law-w_c} presents the dependencies of $\Delta {\rm Prob} \equiv \Delta{\rm Prob}^{[0]}$ on $\lambda^2$ for
the cutoff frequency $\hbar\omega_c = 0.4h_z$.
The dependencies for three values of $\Omega$ exhibit behavior
similar to that presented in Fig.~\ref{DIFGa}. The resonance present for $\hbar\Omega = 4.6h_z$ is still present,
and it is responsible for the deviation from the Boltzmann distribution in the limit of weak coupling,
$\lambda^2 \leq 10^{-4}$. The increase of the system-bath coupling suppresses the resonance,
and the distribution of the diagonal elements of the asymptotic density matrix
approaches the Boltzmann distribution; see inset in Fig.~\ref{law-w_c}.
Therefore, we conclude that all the three conditions stated in Sec.~\ref{condition} [(1), (2), and (3)] are mutually independent and crucial for the existence of the effective Floquet-Gibbs state.
It should be noted here that even in this ideal situation the effective Floquet-Gibbs form presents an approximation and is not
exact as in the case of a time-independent system very weakly coupled to a heat bath.

\section{Conclusion and Discussion}
\label{conclusion}
Dissipation plays a leading role in shaping of
the asymptotic state of a periodically
driven quantum system even in the limit of weak but finite coupling to a heat bath. If certain conditions are met, this state
is characterized by a density matrix of the effective Floquet-Gibbs form, Eq.~(\ref{density}).
These conditions are specified by the relations between characteristic timescales of the three constituents, that are
the system, the bath and the periodic driving field. Namely, (1) the dissipation rate of
the system (controlled by the interaction with the heat bath) must be higher than its heating rate
(controlled by the interaction with the driving field) $\tau_{\rm heat} \gg \tau_{\rm relax}$, (2) the time-dependent part of the system Hamiltonian should commute with itself at different times $[H_{\rm ex}(t_1),H_{\rm ex}(t_2)]=0$, and
(3) the frequency of the driving field should be much larger than the cutoff frequency of the bath spectral density $\omega_c \ll \Omega$ or $[H_{\rm I}, H_{\rm ex}(t)]=0$.
Condition (1) guarantees that the resonance transitions between the energy states
of the effective Floquet Hamiltonian are suppressed while condition (3) insures that
the external driving cannot stimulate excitations inside the bath. Condition (2) is not restrictive and is obeyed in most cases. We validated the theory with the realization of the effective Floquet-Gibbs state
for the case of a non-integrable spin chain and conclusively demonstrated that our above stated conjecture holds.
%We illustrated these conjecture by using a non-integrable spin chain model.

On the way to find the effective
Floquet-Gibbs states we met an intriguing phenomenon that is the existence
of an \emph{effective} temperature different from the actual temperature of the heat bath.
This phenomenon was first observed when studying thermodynamics of an ac-driven dissipative single-particle system,
a quantum nonlinear oscillator \cite{ketzmerick2010statistical}. In this work the appearance of
effective temperature was
related to the existence of the `regular' and `chaotic' Floquet states, defined
in the framework of the semi-classical eigenfunction hypothesis~\cite{berry1977}.
How this could be interpreted in the case of many-body quantum systems and what is the effect of alternative expansion schemes~\cite{Eckardt2015} remain as open issues.

%\bibliography{Augsburg.bib}

\ack
T.S. acknowledges JSPS for financial support (Grant No. 258794). T.S. is supported by Advanced Leading Graduate
Course for Photon Science (ALPS).
JT acknowledges financial support from SMART.
T.M. is supported by JSPS KAKENHI Grant No. 15K17718.
P.H. and S.D. acknowledge support of the Russian Science Foundation (project No. 15-12-20029).
%S. D. and P. H. acknowledge DFG Grants No. HA1517/31-2 and DE1889/1-1.
S.M. was supported by Grants-in-Aid for Scientific Research C (25400391) from MEXT of Japan.
The numerical calculations were supported by the supercomputer center of ISSP of Tokyo University.
T.S., T.M., and S.M. also acknowledge the JSPS Core-to-Core Program: Non-equilibrium dynamics of soft matter and information.

\pagebreak
\appendix

\section{Truncation dependence\label{trunc}}
\begin{figure}[t]
\begin{center}
\includegraphics[width=0.5\textwidth]{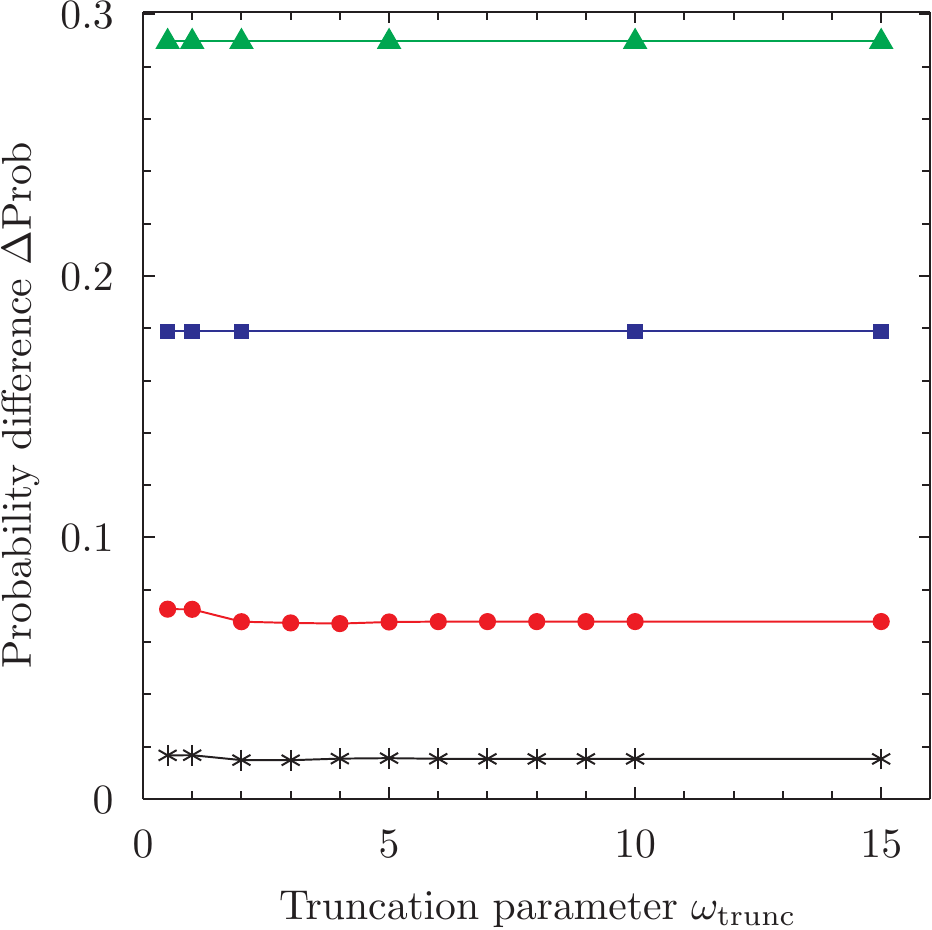}
\end{center}
\caption{Probability difference $\Delta {\rm Prob} \equiv \Delta {\rm Prob}^{[0]}$ vs the truncation frequency $\omega_{\rm trunc}$ at $\hbar \Omega =4.6 h_z$
for sets of system-bath coupling $\lambda^2$ and cutoff frequency $\omega_c$, $(\lambda^2, \hbar\omega_c)=(10^{-2}, 100 h_z)$ ({\color{red}$\bullet$}),
$(\lambda^2, \hbar\omega_c)=(10^{-6}, 100 h_z)$ ({\color{blue}$\subHalmos$}), $(\lambda^2, \hbar\omega_c)=(10^{-2}, 0.4 h_z)$ ($\ast$), and $(\lambda^2, \hbar\omega_c)=(10^{-6}, 0.4 h_z)$ ({\color{green}$\blacktriangle$}).
}
\label{fig1_supp}
\end{figure}

In Fig.~\ref{fig1_supp}, we show the dependence of the probability difference $\Delta{\rm Prob} \equiv \Delta {\rm Prob}^{[0]}$ on the truncation frequency $\omega_{\rm trunc}$. The probability differences are stable against the change of $\omega_{\rm trunc}$. Although setting $\omega_{\rm trunc} = 1$ gives well approximated values of $\Delta{\rm Prob}$, in this work we adopt $\omega_{\rm trunc} = 10$.
\pagebreak
\section{Time-independence of $\Delta {\rm Prob}^{[t]}$\label{timedependence}}
\begin{figure}[t]
\begin{center}
\begin{tabular}{cccc}
&\includegraphics[width=0.45\textwidth]{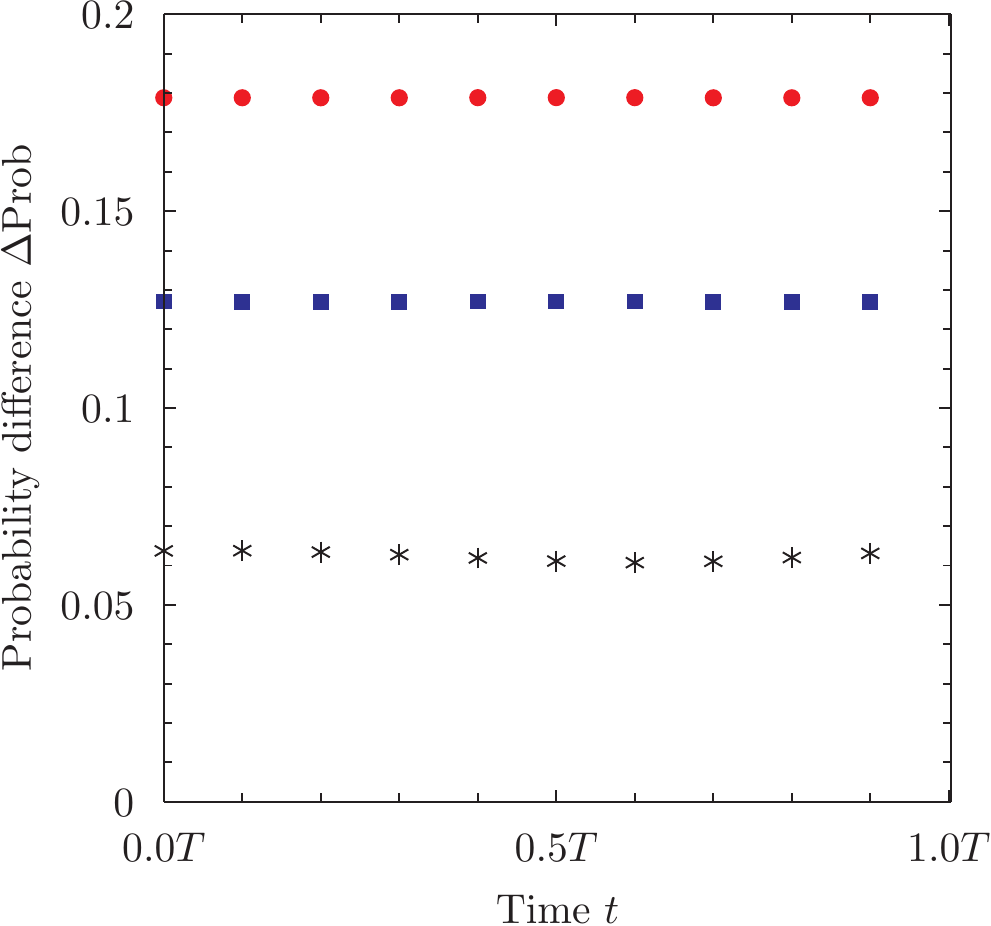}
&&\includegraphics[width=0.45\textwidth]{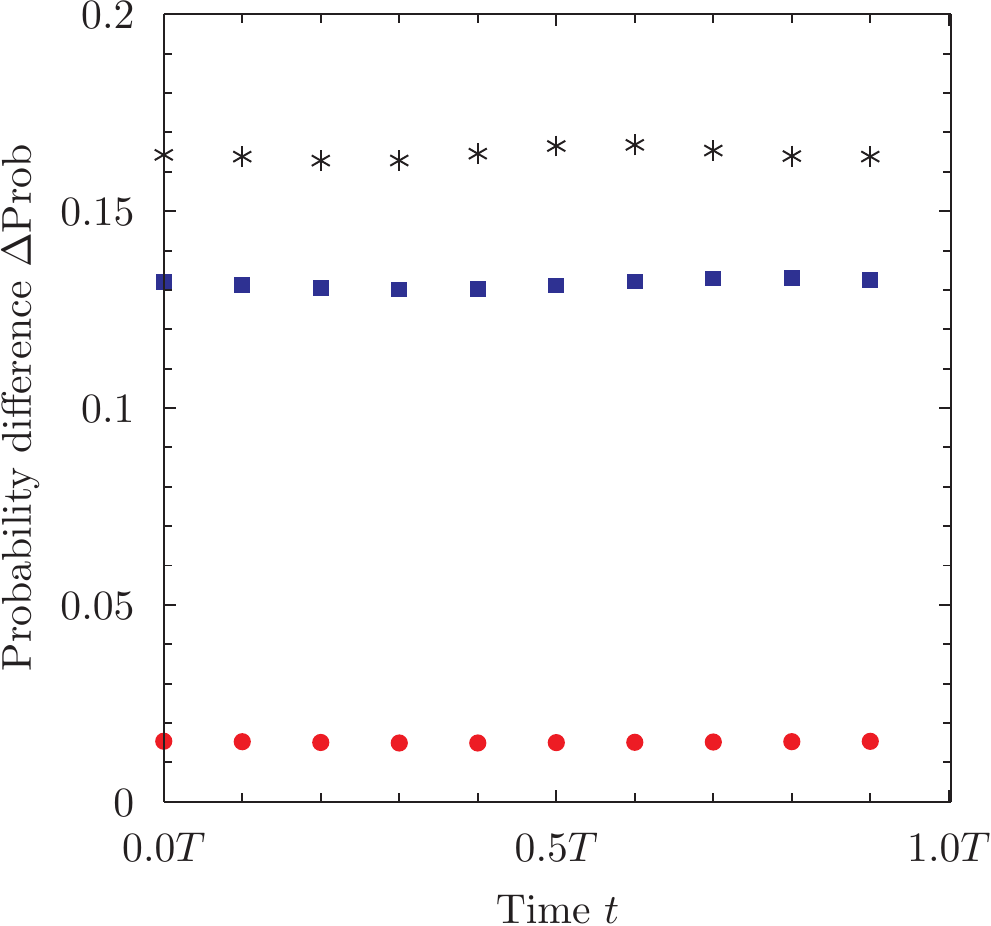}
\end{tabular}
\caption{
Time dependence of probability differences $\Delta{\rm Prob}^{[t]}$ for $\hbar\Omega=4.6 h_z$.
Left figure shows data for $\lambda^2=10^{-6}$ ({\color{red}$\bullet$}), $\lambda^2=2^6*10^{-6}$ ({\color{blue}$\subHalmos$}), and $\lambda^2=2^{14}*10^{-6}$ ($\ast$) at $\hbar \omega_c=100 h_z$.
Right figure shows data for $\hbar \omega_c=0.4 h_z$ ({\color{red}$\bullet$}), $\hbar \omega_c=3.2 h_z$ ({\color{blue}$\subHalmos$}), and $\hbar \omega_c=100 h_z$ ($\ast$) at $\lambda^2=0.01$.
Thus, the probability differences are almost independent  of $t$.
}
\label{fig2_supp}
\end{center}
\end{figure}

In Fig.~\ref{fig2_supp} we show  the time dependence for the $\Delta {\rm Prob}^{[t]}$ plotted in Figs.~4 and 6 in the main text. Since the quantity $\Delta {\rm Prob}^{[t]}$ is time-independent, $\Delta {\rm Prob}^{[0]}$ forms a good representative of the trace distance; i.e., $\Delta {\rm Prob}^{[t]} = \Delta {\rm Prob}^{[0]} \equiv \Delta {\rm Prob}$.

\end{document}